\documentclass[prd,aps,tightenlines,twocolumn,nofootinbib]{revtex4}

\newcommand{\be}{\begin{equation}}
\newcommand{\ee}{\end{equation}}
\newcommand{\bea}{\begin{eqnarray}}
\newcommand{\beas}{\begin{eqnarray*}}
\newcommand{\eea}{\end{eqnarray}}
\newcommand{\eeas}{\end{eqnarray*}} 
\newcommand{\ba}{\begin{array}}
\newcommand{\ea}{\end{array}}

\begin{document}

\title{An Introduction to the Brane World\footnote{Lectures given at the 
 V Mexican Workshop on Gravitation and Mathematical Physics, 
 Morelia, M\'exico. November 24 - 29, 2003}}

\author{Abdel P\'erez-Lorenzana}
\affiliation{Departamento de F\'{\i}sica, Centro de Investigaci\'on y de Estudios 
Avanzados del I.P.N. \\
Apdo. Post. 14-740, 07000, M\'exico, D.F., M\'exico}

\begin{abstract}
The study  of the so called brane world models  has 
introduced completely new ways of looking up on standard problems
in many areas of theoretical physics. 
Inspired in the recent developments of string theory,
the Brane World picture involves 
the introduction  of new extra dimensions beyond the four we see,
which could either be compact or even open (infinite). 
The sole existence of those new dimensions may have non trivial observable 
effects  in  short distance  gravity experiments, 
as well as in our understanding of the cosmology of the early Universe, 
among many other issues.
The goal of the present notes is to provide a short introduction to 
some basic aspects of the  Brane World models.  The discussion
includes models with flat compact extra dimensions, as well as the so called
Randall-Sundrum models.
\end{abstract}

\maketitle

\section{Introduction}

In the course of the last five years there has been  considerable activity in 
the study of models that involve new extra dimensions. The possible existence
of such dimensions got  strong motivation from theories that try to
incorporate gravity and  gauge interactions in a unique scheme,  in
a reliable manner.  The idea  dates back to the 1920's,  to the works of
Kaluza and Klein~\cite{kk} who tried to unify electromagnetism with Einstein
gravity by assuming that the photon originates from the fifth component of the
metric. With the advent of string theory the idea  has gained support
since all versions of string theory are naturally and consistently formulated
only in a  space-time of more than four dimensions  (actually 10D, or 11D for
M-theory). Until recently,  however, it was  conventional to assume that such
extra dimensions were compactified to manifolds  of small radii with sizes  of
the order of the  Planck length,    
$\ell_P\sim ~ 10^{-33}$~cm. or
so, such that they would remain hidden to the experiment. It was only along the
last years of the XX century when people started to ask the question of how
large could these extra dimensions  be without getting into conflict with
observations, and even more interesting, where and how could this extra
dimensions manifest themselves.  The intriguing answer to the first question 
point towards  the possibility that extra dimensions as large as
millimeters~\cite{dvali} could exist  and yet  remain hidden to the
experiments~\cite{expt,giudice,lykken,coll1,sn87,coll2}.  To allow that, however, matter
should be localized to a hypersurface (the brane)  embedded in a higher
dimensional world (the bulk).  
Again, the main motivation for 
these models comes from string theories where the
Horava-Witten solution~\cite{witten} 
of the non-perturbative regime of the $E_8\times E_8$ 
string theory provided one of the first models of this kind.
To answer the second question many
phenomenological studies have been done  in a truly bottom-up approach,  
often based on simplified field theoretical models, 
trying to provide new insights to the possible implications of the fundamental
theory at observable level, although it is unclear whether any of those models
are realized in nature. Nevertheless, they may help to find the way to 
search of extra dimensions, if there is any.

It is fair to say that  similar ideas  were 
proposed on the 80's by several authors~\cite{rubakov}, nevertheless, 
they were missed by some time, until recent developments on string theory, 
basically the raise of M-theory, provided an independent realization 
to such models~\cite{witten,anto,lykk,dbranes}, given them certain credibility. 

It is the goal of the present notes to provide a brief introduction for the
beginner to  general aspects of  theories with extra dimensions. 
Many variants of the very first model
by Arkani-Hammed, Dimopoulos and D'vali~\cite{dvali} have
been proposed over the years, and there is no way we could comment 
all those results in a short review as the present one.
We shall rather concentrate on some of the most general  
characteristics  shared by these models. 
With particular interest we will
address dimensional reduction, which provides the effective four dimensional 
theory on which most calculations are actually done. 
The determination of the effective  gravity coupling, an the also effective 
gravitational potential in four dimensions will be discussed along the notes.
We will also cover some aspects of the cosmology of models in more than four
dimensions. Of particular interest in our discussions 
are the Randall and Sundrum models~\cite{merab,rs1,rs2,nimars} 
for warped backgrounds,
with compact or even infinite extra dimensions.
We will show in detail
how these solutions arise, as well as how gravity behaves in such theories. 
Some further ideas that include: stabilization of the extra dimensions; 
graviton localization at branes; 
and brane cosmology; are also covered. The interested reader that would
like to go beyond the present notes can consult any of the 
excellent reviews that
are now in the literature, some of which are given in reference~\cite{reviews}.

%%%%%%%%%%%%%%%%%%%%%%%%%%%%%%%%%%%%%%%%%%%%%%%%%%%%%%%%%%%%%%%
\section{General Aspects: Flat Extra Dimensions}
%%%%%%%%%%%%%%%%%
\subsection{Planck versus Fundamental Gravity Scales}

The possible existence of more than four dimensions in nature,  even if they
were small, may  not be completely harmless. In principle, they could have
some visible manifestations in our (now effective) four dimensional world. 
To look for such signals, one has first to understand how the effective four
dimensional theory arises from the higher dimensional one. Formally,  this can
be achieve by dimensionally reducing the complete theory, a concept that  we
shall  discuss  further in the following section. One of the first things  we
must notice is that since gravity is a geometric property of the space,  in a
higher dimensional world, where Einstein gravity is assumed to hold, the 
gravitational coupling does not necessarily coincide with the well known Newton
constant $G_N$, which is, nevertheless, the  gravity coupling we do observe. 
To explain this more clearly, let us assume as in
Ref.~\cite{dvali} that there are $\delta$  
extra space-like dimension which are
compactified into circles of the  same radius R 
(so the space is factorized as a ${\cal M}_4\times T^\delta$ manifold).
We will call the fundamental gravity coupling $G_\ast$, and then write down the
higher dimensional gravity action:
 \be 
 S_{grav} = -\frac{1}{16\pi G_\ast} \int \! d^{4+\delta} x
 ~\sqrt{|g_{(4+\delta)}|}~ R_{(4+\delta)}~; 
 \label{sgrav}
 \ee
where $g_{(4+\delta)}$ stands for the metric in the whole $(4+\delta)$D space, 
 \be 
     ds^2 = g_{MN}dx^Mdx^N~,
   \ee
for which we will always use the  $(+,-,-,-,\dots)$ sign convention, and 
$M,N = 0,1,\dots, \delta +3$. 
The above action has to have proper dimensions, 
meaning that the extra length
dimensions that come from the extra volume integration have to be equilibrated
by the dimensions on the gravity coupling. Notice that in natural units,
$c=\hbar =1$;  $S$ has no dimensions. We are also assuming for simplicity
that  $g_{(4+\delta)}$ is being taken dimensionless, so $[R_{(4+\delta)}] =
[Length]^{-2}$ -- or $[Energy]^2$ in natural units --. 

Now, in order to extract the four dimensional gravity action let us  assume
that the extra dimensions are flat, thus, the metric has the form 
   \be 
     ds^2 = g_{\mu\nu}(x) dx^\mu dx^\nu - \delta_{ab}dy^a dy^b,
   \ee
where $g_{\mu\nu}$ gives the four dimensional part of the metric 
which depends only in the four dimensional coordinates $x^\mu$, 
for $\mu=0,1,2,3$; and 
$\delta_{ab}dy^a dy^b$ gives the line element on the torus, 
whose coordinates are parameterized by $y^a$, $a=1,\dots,\delta$.
It is now easy
to see that $\sqrt{|g_{(4+\delta)}|}= \sqrt{|g_{(4)}|}$ and 
$R_{(4+\delta)} = R_{(4)}$, such that one can integrate out the extra
dimensions in Eq.~(\ref{sgrav})  to get the effective action
  \be 
  S_{grav} = -\frac{V_\delta}{16\pi G_\ast} \int \! d^{4}x 
  ~\sqrt{|g_{(4)}|}~ R_{(4)}~; 
  \label{eq4}
  \ee 
where $V_\delta$ stands for the volume  of the extra space. For the torus
$V_\delta=(2\pi R)^\delta$.
Equation (\ref{eq4}) is precisely the 
standard gravity action in 4D if one makes the
identification, 
  \be 
    G_N = G_\ast/V_\delta~.
  \label{add0}
 \ee
Newton constant is therefore given by a volumetric scaling of the truly
fundamental gravity scale.  Thus, $G_N$ is in fact an effective quantity.
Notice that even if $G_\ast$ were a large coupling, one can still 
understand the smallness of $G_N$ via the volumetric suppression. 

To get a more physical meaning of these observations, let us consider a simple
experiment. Let us assume a couple of particles of masses $m_1$ and 
$m_2$, respectively, located on the hypersurface $y^a = 0$, and separated from 
each other  by a  distance $r$. The gravitational flux among both
particles would spread all over  the whole $(4+\delta)$~D space, however, since
the extra dimensions are compact,  the effective strength of the gravity
interaction would have two clear limits: 
(i) If both  test particles are
separated from each other  by a  distance $r \gg R$, 
the torus would effectively 
disappear for the four dimensional observer, the gravitational flux then gets
diluted by the extra volume and the observer would
see the  usual (weak) 4D gravitational potential
 \be 
 U_N(r) = -G_N\frac{m_1 m_2}{r}. \label{ur1}~.
 \ee
(ii) However, if $r \ll R$, the 4D  observer would be able to
feel the presence of the bulk through  
the  flux that goes into the extra space, and 
thus, the  potential between each particle would appear to be stronger:
\be
U_\ast(r)= - G_\ast \frac{m_1 m_2}{ r^{\delta+1}}.
\ee
It is precisely the volumetric factor which does the matching 
of  both regimes of the theory. 
The change in the short distance behavior of the Newton's gravity law
should be observable in the experiments when measuring $U(r)$ for distances below
$R$. Current search for such deviations 
has gone down to 200 microns, so far with no signals of extra
dimensions~\cite{expt}.

We should now recall that the Planck scale, $M_P$,  usually assumed to be the
fundamental energy scale
associated to the scale at which quantum
gravity (or string theory) should make itself manifest, 
is defined in terms of the Newton constant, via
  \be 
  M_P = \left[\frac{\hbar c^5}{8\pi G_N}\right]^{1/2} ~\sim~  
   2.4\times 10^{18}~{\rm GeV}~.
  \ee
In the present picture, it is then clear that $M_P$ is not fundamental anymore. 
The true scale for quantum gravity should be given in terms of $G_\ast$ instead.
We then define the string scale as
  \be 
  M_\ast =   \left[
  \frac{\hbar^{1+\delta} c^{5+\delta}}{8\pi G_\ast}\right]^{1/(2+\delta)}~.
  \ee
Switching to natural units ($c=\hbar =1$) from here on, 
both scales are then related to each other by~\cite{dvali}
 \be
  M_{P}^2 = M^{\delta+2} V_\delta  ~.
  \label{mp}
 \ee
From the particle physics world we already know that there is no evidence of 
quantum gravity (neither supersymmetry, nor  string effects) well up to
energies around few hundred GeV, which says that $M_\ast\geq 1$~TeV. 
If the volume were large enough, 
then the fundamental scale could be as low as the electroweak scale, 
and there would be no hierarchy in the fundamental scales of physics, 
which so far has been considered as a puzzle. 
Of course, the price of solving the hierarchy problem this way 
would be now to explain why the extra dimensions are so large.
Using 
$V\sim R^\delta$ one can reverse above relation and get a feeling of the
possible values of $R$ for a given $M_\ast$. This is done 
just for our  desire of
having the quantum gravity scale as low as possible, although, the
actual value is unknown. As an example, if one takes $M_\ast$ to be 1~TeV; for
$\delta=1$, $R$ turns out to be about the size of the solar system 
($R\sim 10^{11}$~m)!, whereas for $\delta=2$ one gets $R\sim 0.2$~mm, that is
just at the current limit of the experiments. 
More than two extra dimensions are in fact  expected
(strings predicts six more), but in the final theory
those dimensions may turn out to have different sizes, or even geometries. 
More complex scenarios with a hierarchical distributions of the
sizes could be natural. For getting an insight of the  
theory, however, one
usually  relies in toy models with a single  extra dimension, compactified into
circles or orbifolds.

%%%%%%%%%%%%%%%%%%%%
\subsection{Brane World theory prescriptions}

While submillimeter dimensions remain untested for gravity, the particle
physics forces have certainly been accurately measured  up to weak scale
distances (about $10^{-18}$~cm).   Therefore, the matter particles can not
freely propagate in those large extra dimensions,  but must be constrained to
live on  a four dimensional submanifold. Then the  scenario  we have in mind is
one where we live in a four dimensional surface embedded in a higher
dimensional space. Such a surface shall be called a ``brane'' (a short name for
membrane). This picture is similar to the D-brane models~\cite{dbranes},  as in
the Horava-Witten theory~\cite{witten}. We may also imagine our world as a 
domain wall of size $M_\ast^{-1}$  where the particle fields are trapped by
some dynamical mechanism~\cite{dvali}. Such hypersurface or brane would then be
located at an specific point on the extra space, usually, at the fixed points
of the compact manifold. Clearly, such picture breaks translational invariance,
which may be reflected in two ways in the physics of the model, affecting the
flatness of the extra space (which compensates for the  required flatness of
the brane), and introducing a source of violation  of the extra linear
momentum. First point would drive us to the Randall-Sundrum Models, that we
shall discuss latter on. Second point would be a constant issue along our
discussions.

What we have called a brane in our previous paragraph is actually an effective
theory description. We have chosen to think up on them as  
topological defects (domain walls) of almost zero width, 
which could have fields localized to  its surface.
String theory D-branes  (Dirichlet banes)
are, however,  surfaces where open string can end on. 
Open strings  give rise to all kinds of fields localized 
to the brane, including gauge fields. In the supergravity 
approximation these D-branes will also appear as solitons of the 
supergravity equations of motion. 
In our approach we shall care little about  
where these branes come from, and rather simply assume
there is some consistent high-energy theory,  that 
would give rise to these objects, and which should appear at the fundamental
scale $M_\ast$. Thus the natural cutoff of our models would always be given by
the quantum gravity scale.

D-branes are usually characterized by the number of spatial dimensions on the
surface. Hence, a p-brane is described by a flat space time with 
p space-like  and one  time-like coordinates. 
Unless otherwise stated, we will always  work with models of 3-branes.
We need to be able to describe theories that live either in the brane 
(as the Standard Model) or in the bulk (like gravity), 
as well as the  interactions among these two theories.
For doing so we use the following field theory prescriptions: 
\begin{itemize}
\item[(i)] Bulk theories are 
 described by the higher dimensional action, defined in terms of a 
Lagrangian density of $\phi(x,y)$ fields valued  
on all the space-time coordinates of the  bulk 
 \be 
    S_{\rm bulk}[\phi] = \int \! d^4x\, d^\delta y~
    \sqrt{|g_{(4+\delta)}|}{\cal L}(\phi(x,y))~,
   \label{sbulk}
 \ee
where, as before, $x$ stands for the (3+1) coordinates of the brane and 
$y$ for the $\delta$ extra dimensions. 
\item[(ii)] Brane theories are described by the (3+1)D action of the brane
fields, $\varphi(x)$, which is
naturally promoted into a  higher dimensional expression by the use of a 
delta density:
 \be 
    S_{\rm brane}[\varphi] = \int \! d^4x\, d^\delta y~
    \sqrt{|g_{(4)}|}{\cal L}(\varphi(x))~
    \delta^\delta({\vec y}-{\vec y}_0)~,
   \label{sbrane}
 \ee
where we have taken the brane to be 
located at the position ${\vec y}={\vec y}_0$
along the extra dimensions, and the metric $g_{(4)}$ stands for the 4D induced
metric on the brane.
\item[(iii)] Finally, the action may contain terms that couple bulk to brane
fields. Last are localized on the space, thus, 
it is natural that a delta density would be involved 
in   such terms, say for instance
 \bea 
    &\propto& \int \! d^4x\, d^\delta y~ 
    \sqrt{|g_{(4+\delta)}|}~ \phi^2(x,y)\, \varphi(x)~
    \delta^\delta({\vec y}-{\vec y}_0)
    \nonumber \\
    &=&
    \int \! d^4x ~  \sqrt{|g_{(4)}|}\phi^2(x,0) \varphi(x)~.
   \label{bb0}
 \eea
 
\end{itemize}

%%%%%%%%%%%%%%%%%%%%
\subsection{Dimensional reduction: Kaluza-Klein Decomposition}

The presence of delta functions in the previous actions does not allow for a
transparent interpretation, nor for an easy reading out 
of the  theory  dynamics.  
When they are present it is more useful to work in the effective four
dimensional theory which is  obtained after integrating over the extra
dimensions. This procedure is generically called dimensional reduction.
It also helps to identify the low energy limit of the theory (where the extra
dimensions are not visible).

To get an insight of what the effective 4D theory looks like, 
let us consider a simplified five dimensional toy model where the
fifth dimension has been compactified on a circle of radius $R$. 
The generalization of these results to more dimensions would be 
straightforward.
Let $\phi$ be  a  bulk scalar field  for which the action
on flat space time has the form 
 \be  
 S[\phi] = \frac{1}{2}\int\! d^4x\ dy\  
 \left(\partial^A\phi\partial_A\phi  - m^2 \phi^2\right) ;
 \ee
where now $A=1,\dots,5$, and $y$ denotes the fifth dimension. 
The compactness of the internal manifold is reflected in the 
periodicity of the field,  
$\phi(y) = \phi(y+2\pi R)$, which allows for a  
Fourier expansion of the field as 
 \bea
 &\phi(x,y)& = \frac{1}{\sqrt{2\pi R}}\phi_0(x) 
  +   \sum_{n=1}^\infty \frac{1}{\sqrt{\pi R}} 
  \left[ \phi_n(x)\cos\left(\frac{ny}{ R}\right) \right.\nonumber    \\
  && + \left. \hat\phi_n(x)\sin\left(\frac{ny}{R}\right)\right]. 
 \label{kkexp}
 \eea 
The very first term, $\phi_0$, 
with no dependence on the fifth dimension is 
usually referred as the zero mode. Other Fourier modes,
$\phi_n$ and $\hat\phi_n$; are called
the excited or Kaluza-Klein (KK) modes.
Notice the different normalization on all the excited modes, $\phi_n$ 
and $\hat\phi_n$, with respect to the zero mode.  

By introducing
last expansion into the action and integrating over the extra
dimension one gets
 \bea 
 S[\phi] &=& \sum_{n=0}^\infty \frac{1}{2}\int\! d^4x 
 \left(\partial^\mu\phi_n\partial_\mu \phi_n - m_n^2 \phi_n^2\right)
 \nonumber \\
 &&+  \sum_{n=1}^\infty \frac{1}{2}\int\! d^4x 
 \left(\partial^\mu\hat\phi_n\partial_\mu\hat\phi_n - 
 m_n^2\hat\phi_n^2\right),
 \eea
where the KK mass is given as $m_n^2 = m^2 + \frac{n^2}{R^2}$. Therefore,
in the effective theory, the higher dimensional field appears as an
infinite tower of  fields with masses $m_n$.

Notice that all excited  modes are fields with the same spin, and
quantum numbers as $\phi$.  But they differ in the KK number $n$,
which is also associated with the fifth component of the momentum. 
From a formal point of view, KK modes are only a manifestation of the
discretization of the  extra momentum of the
particle. We would see particles with different higher dimensional momentum as 
having different masses. 
This can be also understood from the higher dimensional 
invariant $p^A p_A  = m^2$, which can be rewritten as the effective four
dimensional squared momentum invariant 
$p^\mu p_\mu = m^2 + {{\vec p}_\bot}\,^2$, 
where  ${\vec p}_\bot$ stands for the extra momentum components.

Dimensionally reducing any higher dimensional field
theory would indeed give a similar spectrum for each particle. 
For $m=0$ it is clear that
for energies below $\frac{1}{R}$ only the massless zero mode
will be kinematically accessible, making the theory looking four dimensional.
The appreciation of the impact of KK excitations
thus depends on the relevant energy of the experiment,
and on the compactification scale $\frac{1}{R}$:
\begin{itemize}
\item[(i)]
For energies $E\ll \frac{1}{R}$ 
physics would  behave purely four dimensional.
\item[(ii)] At larger energies, $\frac{1}{R}<E<M_\ast$,
 or equivalently as we do measurements at shorter
distances, a large number 
of  KK excitations, $\sim (ER)^\delta$, becomes kinematically
accessible, and their contributions relevant for the physics. 
Therefore, right  above the threshold of the first excited level, 
the manifestation of the KK modes will start evidencing  the higher
dimensional nature of the theory. 
\item[(iii)]
At energies above $M_\ast$, however,  our effective approach has to be 
replaced by the use  of the fundamental theory that describes 
quantum gravity phenomena.
\end{itemize}

Furthermore, notice that the five dimensional scalar field $\phi$ 
we considered before has mass 
dimension $\frac{3}{2}$, in natural units. 
This can be easily seeing from the kinetic part 
of the Lagrangian, which involves two
partial derivatives with mass dimension one each, 
and the fact that the action should be dimensionless. 
In contrast, by similar arguments, all excited modes have mass 
dimension one, which is consistent with the KK expansion (\ref{kkexp}). 
In general for $\delta$ extra dimensions we  get the mass dimension for 
an arbitrary field to be $[\phi] = d_4 + \frac{\delta}{2}$, 
where $d_4$ is the natural mass dimension of  $\phi$
in four dimensions. 
Because this change on the dimensionality of
$\phi$, most interaction terms on the Lagrangian 
(apart from the mass term) 
would  all have dimensionful couplings.
To keep them dimensionless a mass parameter should be  introduced to
correct the dimensions.
It is common to use as the natural choice for this parameter
the cut-off of the theory, $M_\ast$. 
For instance, let us consider the quartic
couplings of $\phi$ in 5D. Since all potential terms should be of dimension
five, we should write down $\frac{\lambda}{M_\ast} \phi^4$.  After
integrating the fifth dimension, this operator will generate quartic 
couplings among all KK modes. Four normalization factors containing
$1/\sqrt{R}$ appear in the expansion of $\phi^4$. Two of them will be 
removed by the integration, thus,  we are left with the  effective
coupling $\lambda/MR$.  By introducing Eq. (\ref{mp}) we  observe that the
effective couplings have the form
 \be 
 \lambda\left( \frac{M}{M_{P}}\right)^2 \phi_k\phi_l\phi_m\phi_{k+l+m}.
 \ee
where the indices are arranged to respect the conservation of the fifth
momentum.  From the last expression we conclude that in the low energy theory
($E<M_\ast$),  even at the zero mode level,  the effective coupling appears
suppressed respect to the bulk theory.  Therefore, the  effective four
dimensional theory would be  weaker interacting compared to the bulk
theory.   Let us recall that  same  happened to gravity on the bulk, 
where  the
coupling constant is stronger than the effective 4D coupling, due to the 
volume suppression given in Eq.~(\ref{add0}), or equivalently in
Eq.~(\ref{mp}). Similar arguments apply in general for the brane-bulk couplings.
We shall use these facts when considering more carefully gravity interactions 
for test particles on the brane, which we shall do along the next section.

Different compactifications would lead to different mode expansions. 
Eq. (\ref{kkexp}), would had to be chosen  accordingly to the geometry of  the
extra space, by typically using  
wave functions for free particles on such space as the basis for the expansion.
Extra boundary conditions associated to specific topological properties of the
compact space may also help for a proper selection of the basis. 
A useful example is the one dimensional orbifold, $U(1)/Z_2$, 
which is built out of the circle, 
by identifying the opposite points around zero. The operations can be
see as reducing the interval of the original circle to $[0,\pi]$ only. 
Operatively, this is done by requiring the theory to be invariant under the 
extra parity symmetry $Z_2: y\rightarrow -y$. Under this symmetries all fields 
should pick up a specific parity. Even (ood) fields would then be expanded into 
only cosine (sine)  modes.
Thus, odd fields do not appear at the 
zero mode level  of the theory, which also means that the
orbifolding  projects out half  of the modes on the KK expansion.

%%%%%%%%%%%%%%%%%%
\subsection{Graviton couplings and the effective gravity
interaction law.}

One of the first physical examples of a brane-bulk interaction one may be 
interested in analyzing with some  care is the effective 
gravitational coupling of particles  located at the brane, 
which needs to understand the way gravitons couple to brane fields.
The problem has been extendedly discussed  by Giudice, Ratazzi and 
Wells~\cite{giudice}
and independently by Han, Lykken and Zhang~\cite{lykken} assuming a flat bulk. 
Here we summarize some of the main points.
We start from the action that describes a particle on the brane
 \be 
 S = \int \!d^4x \sqrt{|g(y^a=0)|}~{\cal L} ~,
 \ee
where the induced metric $g(y^a=0)$ 
now includes  small metric fluctuations 
$h_{MN}$ over flat space, which are also called the graviton, such that 
 \be 
 g_{MN} = \eta_{MN} + \frac{1}{2M_\ast^{\delta/2+1}} h_{MN}. 
 \ee
The source of those fluctuations are of course the energy on the brane, i.e.,
the matter 
energy momentum tensor $\sqrt{g}~ T^{\mu\nu} = \delta S/\delta g_{\mu\nu}$ 
that enters on the RHS of Einstein equations:
\[R_{MN} - \frac{1}{2} R_{(4+\delta)}g_{MN} = -\frac{1}{M_\ast^{2+\delta}}
T_{\mu\nu}\eta^{\mu}_M \eta^\nu_N \delta^{(\delta)}(y)~.\] 
The effective coupling, at first order in $h$ of matter to graviton field is
then described by the action
 \be 
 S_{int} = \int\!d^4x \frac{h_{\mu\nu}}{M_\ast^{\delta/2 +1}}T^{\mu\nu} 
 \label{gravt}
 \ee
 
It is clear  from the effective four dimensional
point of view, that the fluctuations $h_{MN}$ would have different 4D Lorentz
components. (i) $h_{\mu\nu}$ clearly contains a 4D Lorentz tensor, the true 
four dimensional graviton. (ii) $h_{a\mu}$ behaves as a vector, the
graviphotons. (iii) Finally, $h_{ab}$  behaves as a group of scalars
(graviscalar fields), one of which corresponds to the partial trace of $h$ 
(${h^a}_a$) that later we will call the radion field.
To count the number of degrees of freedom in $h_{MN}$ 
we should first note that $h$ is $n\times n$ a symmetric tensor, for
$n=4+\delta$. 
Next,  general coordinate invariance of general relativity can be translated
into $2n$ independent gauge fixing conditions, half usually chosen as the
harmonic gauge  $\partial_M h_N^M = \frac{1}{2} \partial_N h^M_M$. 
In total there are $n(n-3)/2$ independent degrees of freedom. Clearly,
for $n=4$ one has the usual two helicity states of a 
massless spin two particle. 

All those effective fields would of course  have a KK decomposition,
 \be 
 h_{MN}(x,y) = \sum_{\vec{n}} \frac{h_{MN}^{(\vec{n})}(x)}{\sqrt{V_\delta}}~
 e^{i\vec{n}\cdot\vec{y}/R}~.
 \ee
Here $\vec{n} = (n_1,\dots,n_\delta)$, with all $n_a$ integer numbers.
Once we insert back the above expansion into $S_{int}$ it is not hard to see
that the volume suppression will exchange  the  $M_\ast$ by an $M_P$ 
suppression 
in the effective interaction Lagrangian of a single KK mode.
Therefore, all modes couple with the  standard gravity strength.
Briefly, only the 4D gravitons, $G_{\mu\nu}$, 
and the radion field, $\sigma$, get couple at
first order level to the brane energy momentum tensor~\cite{giudice,lykken}
 \be 
 {\cal L} = -\frac{1}{M_P}\sum_{\vec{n}} \left[G^{(\vec{n})\mu\nu} - 
 \frac{\kappa}{3}\sigma^{(\vec{n})}\eta^{\mu\nu}\right] T_{\mu\nu}.
 \label{leff0}
 \ee
Here $\kappa$ is a parameter of order one. 
Notice that $G^{(0)\mu\nu}$ is
massless since the higher dimensional graviton $h_{MN}$ has no mass itself. 
That is the source of long range four dimensional gravity interactions. 
It is worth saying that on the contrary $\sigma^{(0)}$ 
should not be  massless, otherwise it should  violate 
the equivalence principle,
since it would mean a scalar (gravitational) interaction of  long range too. 
$\sigma^{(0)}$ should get a mass
from the stabilization mechanism that keeps the extra volume finite. 
We shall come back to this problem latter on.
From supernova constrains such a mass
should be larger than $10^{-3}$ eV~\cite{sn87}.  

Above Lagrangian runs over all KK levels, 
meaning that brane particles can release
any kind of KK gravitons into the bulk. KK index $\vec n$ is also the extra
component of the momentum, so they go out of the brane taking its energy away, 
in a clear violation of the 4D conservation of energy.
This could appear in  future high energy 
collider experiments, for instance, as missing energy~\cite{coll1,coll2}. 

We started the section asking for the actual form of the effective gravitation
interaction among particles on the brane. Now that we know how gravitons couple
to  brane matter 
we can use this effective field theory point of view to calculate 
what the effective gravitational interaction law should be. KK gravitons are
indeed massive, thus, the interaction mediated by them are of short range. 
More precisely, each KK mode contribute to the gravitational potential among
two test particles of masses $m_1$ and $m_2$ located on the brane, separated
by a distance $r$, with a Yukawa potential  
 \be 
 \Delta_{\vec{n}}U(r) \simeq -G_N \frac{m_1 m_2}{r} e^{-m_{\vec{n}} r}
  = U_N(r) e^{-m_{\vec{n}} r}~.
 \label{yukp}
 \ee
Total contribution of all KK modes, the sum over all KK masses
$m_{\vec{n}}^2 = \vec{n}^2/R$, can be estimated 
in the  continuum  limit, to get 
 \be 
 U_T(r) \simeq - G_N V_\delta (\delta -1)!~ \frac{m_1 m_2}{r^\delta+1}
 \simeq U_\ast(r)~.
 \ee
Experimentally, however, for $r$ just below $R$ only the very first excited
modes would be relevant, and so, 
the potential one should see in short distance
tests of Newton's law~\cite{expt} should rather be of the form 
 \be 
 U(r)\simeq U_N(r) \left( 1+ \alpha e^{-r/R} \right)~.
 \ee
where $\alpha$ stands to account for the multiplicity of the very first excited 
level. 

%%%%%%%%%%%%%%%%%%%%%%%%%%%%%%%%%%%%%%%%%%%%%%%
\section{Cosmology in models with flat extra dimensions}
%%%%%%%%%%%%%%%%

\subsection{Limits on  Reheating Temperature due to  Graviton emission}

Graviton production by brane processes may not be such a harmless phenomena. It
may rather  posses strong constraints on the theory when considering that the
early Universe was an important resource of energy, which in the present
picture resides completely  on the brane. How much of this energy could have
gone into the bulk without affecting cosmological evolution? For large extra
dimensions, the splitting among two excited modes is pretty small, $1/R$. For
$\delta=2$ and $M_\ast$ at TeV scale this means a splitting of just about
$10^{-3}$~eV!. For a process  where the center mass energy is E, up to 
 $N= (ER)^\delta$ KK modes would be  kinematically accessible. During Big Bang
Nucleosynthesis (BBN), for instance, where $E$ was about few MeV,  this already
means more than $10^{18}$ modes.  So many modes may be troublesome, and one has
to  ask the question how hot the Universe could go without loosing too much
energy. By looking at the effective Lagrangian in Eq.~(\ref{leff0}), one can
immediately notice that the graviton creation rate, per unit time and volume, 
from  brane thermal processes at temperature $T$ goes as 
$$ \sigma_{total} = \frac{(TR)^\delta}{M_P^2} =
\frac{T^\delta}{M_\ast^{\delta+2}}. $$
The standard Universe evolution would be conserved, as far as the total number
density of  produced 
KK gravitons, $n_g$,remains small when compared to 
photon number density, $n_\gamma$. This is a
sufficient condition that  can be translated into a bound for the
reheating energy, since as hotter the media as more gravitons can be excited.
It is not hard to see that this condition implies~\cite{dvali} 
 \be 
 \frac{n_g}{n_\gamma} \approx \frac{T^{\delta+1}M_P}{M_\ast^{\delta +2}} <1~.
 \ee
Equivalently, the maximal temperature our Universe could reach with out
producing to many gravitons must satisfy
 \be 
 T_r^{\delta +1}< \frac{M_\ast^{\delta+2}}{M_P}~.
 \label{trgt}
 \ee
To give numbers consider for instance $M_\ast = 10$~TeV and $\delta=2$, 
which means  $T_r< 100 MeV$, just about to what is needed to have 
BBN working. The brane
Universe with large extra dimensions is then rather cold. This would be
reflected in some difficulties for those models trying to implement 
baryogenesis or leptogenesis based in electroweak energy physics or above.
 
As a complementary note, we should mention that since thermal graviton 
emission is not restricted to early Universe, one can expect this to be
happening in any other environment. We have already mention colliders as an
example. But even the hot astrophysical objects can be  gravitons sources. 
Gravitons emitted by stellar objects take away energy, which
contribute to cooling down the star. The stringent bounds on $M_\ast$ actually come
from the study of this process~\cite{sn87}.

%%%%%%%%%%%%%%%%
\subsection{Dimensional reduction and the radion field}

We are now ready to postulate the model that sho\-uld describe the 
brane Universe evolution where
4D Friedmann-Robertson-Walker model must now be obtained as 
the effective zero mode limit, after dimensional reduction.
To simplify the discussion we will  assume  the extra space compactified 
into an orbifolded torus, $T^\delta/Z_2$, such that the extra coordinates 
$y^a$ take values in the interval $[0,1]$, 
and the theory is invariant under the
mapping $Z_2:\vec{y}\rightarrow -\vec{y}$. 
As before the brane should be located at $y^a=0$.
Consider  the metric in  $4+ \delta$ dimensions parameterized by 
 \be
 ds^2 = g_{AB}dx^A dx^B =  g_{\mu\nu}dx^\mu dx^\nu - h_{ab} dy^a dy^b\,,
 \ee
Note that since we have taken $y^a$ to be dimensionless,  $h_{ab}$ has
length dimension two. 
Also note that we are not considering in our parameterization 
the presence of  vector-like connection  $A_\mu^a$ pieces, 
which are common in Kaluza Klein theories. 
This is  
because we would be only interested in the zero mode part of the metric, and 
$A_\mu^a$ vanishes at this level since it is odd under the $Z_2$ parity
transformation.

Next, let us reduce the Einstein-Hilbert action
 \be
 S=
 \frac{1}{2k_{\ast}^2}\, \int\!d^4x\, d^\delta y\, 
 \sqrt{|g_{(4+\delta)}|}\, R_{(4+\delta)} 
 \ee
to four dimensions,  considering only the zero mode level.  
Here, $1/k_{\ast}^2 = M_{\ast}^{2+\delta}$. One then obtains
 \bea
 S&=&
 \frac{1}{2k^2}\, \int\!d^4x\,
 \sqrt{-g_{(4)}}\, \frac{\sqrt{h}}{V_\delta}\, 
 \left \{ R_{(4)} - \frac{1}{4} \partial_\mu h^{ab}\, \partial^\mu h_{ab} 
  \right.
 \nonumber \\
 && \left. -\frac{1}{4}h^{ab}\partial_\mu h_{ab}\cdot h^{cd}\partial^\mu h_{cd}
 \right \}\,,
 \label{rg1}
 \eea
where   $1/k^2 = M_{p}^2$, and 
$V_\delta$ stands for the 
stable volume of the extra space that correspond to Eq.~(\ref{mp}).

In order to obtain the $4$ dimensional scalar curvature term in 
a canonical  form, we have to perform a conformal transformation 
on the metric,
 \be
 g_{\mu\nu}\rightarrow e^{2\varphi}g_{\mu\nu}\,,
 \label{conformal}
 \ee
designed to cancel the extra $\sqrt{h}/V_\delta$ coefficient of $R_{(4)}$ in
Eq.~(\ref{rg1}). We take $\varphi$ such that
 \be
 e^{2\varphi}\, \sqrt{h}/V_\delta = 1\,.
 \label{ephih}
 \ee
The action in Eq.~(\ref{rg1}) is then transformed into
 \bea
 S&=&
 \frac{1}{2k^2}\, \int\!d^4x\,\sqrt{-g_{(4)}}\, 
 \left \{  R_{(4)} - \frac{1}{4} \partial_\mu h^{ab}\, \partial^\mu h_{ab}
 \right.
 \nonumber \\
 && 
 \left. 
 + \frac{1}{8}h^{ab}\partial_\mu h_{ab}\cdot h^{cd}\partial^\mu h_{cd}
 \right \}\,.
 \label{rg2}
 \eea
Next, for the four dimensional part of the metric, $g_{\mu\nu}$, we can now 
assume the standard  Friedmann-Robertson-Walker (FRW) 
metric with a flat geometry,
i.e.  
 \be
    g_{\mu\nu} = diag( 1, -a(t),-a(t), -a(t))\,,
 \label{frw}
 \ee
for an isotropic and homogeneous (brane) Universe, 
whereas we  consider a diagonal form for the $h$ part of the metric:
 \be
 h_{ab} = b(t)^2\delta_{ab}
\label{hij}
 \ee
Obviously, the physical volume of the
extra space is dynamical, and  given as
 \[
 vol_{\rm phys}=\sqrt{h}= b^\delta(t)~.
 \]
If the bulk is stable, meaning a constant $b$, 
the physical size of the extra dimension is  given by
the identification  $b_0= R$.    
This turns out to be the stabilized condition,
when one assumes the volume to have some  dynamics,
which should be reached at some given finite time $t$.

The action can be simplified by defining the radion field
 \be 
 \sigma(t) = M_P\sqrt{\frac{\delta(\delta+2)}{2}}
 \ln\left(\frac{b(t)}{R}\right)~.
 \ee
This has a straightforward 
physical interpretation, 
it is related to the variation of the physical size of the 
volume. Notice that  the radion is set 
to be zero when the stabilized volume is reached.
In this terms  one gets the effective action
 \be 
 S =  \int\!d^4x \frac{\sqrt{-g_{(4)}}}{2k^2}\,  R_{(4)} 
  + \, \int\!d^4x \frac{\sqrt{-g_{(4)}}}{2}\,\left(\partial^\mu \sigma\right)
  \left(\partial_\mu \sigma\right)
 \ee
The very first term corresponds precisely to the 4D gravity action of the 
FRW model. On the other hand, last term can be identified as the action of a
running mode. It is unstable under perturbations, which means that any small
perturbation  on the radion field can make the volume of the extra space expand
or contract without control.
This is what is called the radion stabilization 
problem, and it is a particular case of the more
general moduli problem inherited from string theory. 
Understanding the stability of the volume of the compact space 
can be seeing as finding the mechanism that 
provides the force that keeps the radion fixed at its zero value.
Thus, one has to find the potential $\sigma$ which provides
such a force. 
The  origin of this potential is largely unknown so far, although some ideas
can be found in the literature, see for
instance~\cite{kaluza-klein,tsujikawa}. 
We will not comment on this in here, 
but rather assume that such a potential should exist.
As we will discuss later on,  the detailed form of
$U(\sigma)$  may also be  important
to  understand the dynamics of the radions during and after inflation.
As a parenthetical note, the physical mass of the radion is actually 
related to the second derivative of the potential at minimum, and certainly, to
avoid violation of the equivalence principle, it should be larger than about
$10^{-3}$~eV.

As already mentioned, radion couples to matter fields. 
This is regardless whether they are brane or bulk fields. 
After dimensional reducing the action in Eq.~(\ref{sbulk}),
and including  
the conformal transformation that we performed on the metric, 
we get for the  scalar field the effective action at the
zero mode level
\be
 S[\phi]=\int\! d^4x\,\sqrt{-g_{(4)}}
  \left[\frac{1}{2}\, \partial^\mu\phi\, \partial_\mu\phi
   - e^{-\alpha \sigma / M_{p} }V(\phi)\right]\,,
 \label{sphi}
 \ee
where the coupling constant is given by
$\alpha = \sqrt{2/\delta(\delta+2)}$. The case of a brane scalar field turns out to have
the same functional form for the effective action as above.

%%%%%%%%%%%%%%%%%%%
\subsection{Inflation}

It is still possible that, due to some dynamical mechanism,  the extra
dimension gets stabilized long before the Universe exited from inflation,  as
in some scenarios in  Kaluza-Klein (KK) theories, where the stabilization
potential is generated by the Casimir force~\cite{kaluza-klein,tsujikawa}.
Other possible sources for this stabilizing potential could be present  in
brane-bulk theories; for instance, the formation of the brane at very early
times may give rise to vacuum energy that plays a role in eventually 
stabilizing the extra dimension. 
Let us for the moment consider an stable bulk  ($b=R$), and then 
address the problem of brane cosmological evolution. 
It is clear from the results on previous section, by taking $\sigma=0$, 
that the brane cosmological theory  do behaves  as four dimensional.
The usual FRW model is therefore a good set up to analyze  cosmology on the
brane. During inflation period Hubble expansion is given as usual by 
 \be 
 H\sim \sqrt{\frac{V(\phi)}{3 M_{P}^2}}~,
 \label{eq2}
 \ee
with $V(\phi)$ the potential of the slow rolling inflaton. 
A brane inflaton, however, is troublesome~\cite{inflation1}.  
Consider for instance a typical chaotic inflation scenario~\cite{chaotic}, 
where the potential is simply given by $V(\phi) = \frac{1}{2} m^2 \phi^2$.
If the highest scale in the theory is $M_\ast$, during inflation, 
a brane  inflaton potential  can not get values larger than $M_\ast^4$, 
regardless of the number of extra dimensions. Just as in the usual 4D theories
where the scale of the 
potential is not supposed to be larger than Planck scale.
Next, since successful inflation (the slow roll condition) requires
that the inflaton mass be less than the Hubble parameter, 
we have the inequality  
 \be 
 m\leq H\leq M_\ast^2/M_{P}.
 \label{suppres}
 \ee
For $M_\ast\sim 1$ TeV, one then gets the bound  $ m\leq 10^{-3}$ eV, which 
is a severe fine tuning constraint on the parameters of the theory.
Furthermore, such a light inflaton would certainly face troubles for reheating,
since it would only decay into photons.
The inflaton is believed to be chargeless, thus, such a 
decay can only occur via suppressed loop processes.
Above constraint further implies that  
inflation occurs on a time scale $H^{-1}$ much grater than $M_\ast^{-1}$. 
As emphasized by Kaloper and Linde~\cite{inflation1},
this is conceptually very problematic since it requires that 
the Universe should be large and homogeneous
enough from the very beginning so as to survive the large period of
time from $t=M_\ast^{-1}$ to $t=H^{-1}$.

Moreover for chaotic inflation one gets a tiny contribution to 
density perturbations
 \be 
 \frac{\delta\rho}{\rho}\sim 50 \frac{m}{M_{P}}\leq 10^{-31}.
 \ee
The situation does not improve for more elaborated models.  
For the case where $\lambda\phi^4$ term dominates the density, for instance, 
one gets the same  fine tuning condition than in four dimensions,
 $\frac{\delta\rho}{\rho}\sim \lambda^{1/2}$.
Assuming Hybrid inflation~\cite{hybrid}, with the potential
$V(\phi,\sigma)=
 \frac{1}{4\lambda} \left(M^2 - {\lambda} \sigma^2\right)^2 + 
 \frac{1}{2}m^2 \phi^2 + {g^2} \phi^2 \sigma^2 $,
does not help either~\cite{inflation1}, 
since it needs either a value of $m$ six orders of
magnitude smaller or a strong fine tuning on the parameters, to match
the COBE result $\frac{\delta\rho}{\rho}\sim 10^{-5}$. 
Certainly, the problem would be relaxed if the fundamental scale $M_\ast$ were
much larger than a few TeV, nevertheless, this means a shorter radii and 
most of the phenomenological interest on the model would also be gone.

A simple way to solve this problem could be assuming that the inflaton is the
zero mode of a bulk scalar field~\cite{inf5d}. 
As such, its effective potential energy is
enhanced by the volume of the extra space, 
which allows to have larger densities
contributing to Hubble expansion. 
Indeed, now one has
$$ V_{eff}(\phi_0) = V_\delta~ V_{bulk}(\phi_0)\leq M_\ast^2 M_P;$$
where the RHS comes from the natural upper bound 
$V_{bulk}(\phi_0)<M_\ast^{4+\delta}$. This immediately means that not
stringent bound exists for Hubble, $H\leq M_\ast$, and non superlight inflaton
is required. This
also keeps the  explanation of the flatness and horizon problems as
usual, since  now, 
the time for inflation could be as short as in the standard theory.
Hybrid inflation model now accommodates a nice prediction for density
perturbations~\cite{inf5d},
\be 
  \frac{\delta\rho}{\rho} \sim 
  \left(\frac{g}{2 \lambda^{3/2}}\right)\ \frac{M^3}{m_0^2 M_{P}}~,
 \label{drhyb}
 \ee
which can easily give COBE normalization.
Reheating would now be produced by the decay of the inflaton into brane standard
fields. The effective brane-bulk coupling has a  Planck suppression on it, which
amounts for a low reheating temperature, that nevertheless comes out to be just
right to allow for a successful BBN process. To give numbers, let us 
consider   $T_R \sim 0.1\sqrt{\Gamma_\phi M_{P\ell}}$ 
and a typical rate for the decay into
higgses $\Gamma_\phi\sim M_\ast^4/ (32 \pi M_{P}^2 m_\phi)$, and use 
$m_\phi$
around $0.1 M_\ast$, with $M_\ast\sim 100TeV$, one then gets  
$T_R \sim 100$~MeV. It is worth saying that this number is well within the
constraints  due to graviton thermal production~(\ref{trgt}) discussed
previously.

It is worth noticing that usually the needed number of 
e-foldings is much less than $60$ if the scale of inflation is low.
Especially if the scale of inflation is as low as 
$H_{inf}\approx M_{\ast}\sim$~TeV, it was already shown in 
extra dimensional models~\cite{Lorenzana1,Green} 
that the number of e-foldings required for 
structure formation would be $43$, provided the universe reheats 
by $T_{rh}\sim 10-100$~MeV. Therefore it is only the last $43$ e-foldings of
inflation which are important for the purpose of density perturbations.

As an alternative  way to solve inflaton problems, 
Arkani-Hamed {\it et al}~\cite{arkani-hamed}
proposed a scenario  where it was assumed that inflation occurs  before
the stabilization of the internal dimensions. 
With  the dilaton field  playing the role of the inflaton field, 
they argued that early inflation,
when the internal dimensions are small, can   overcome the  complications we
mention at the beginning of present section. 
However, one can
not allow the extra dimension to grow considerably during inflation since
large changes on the internal size will significantly affect
the scale invariance of the density
perturbations,
The radius of the extra dimension must remain
essentially static while
the Universe expands (slow rolling), this 
needs a quite flat  potential
which  may have trouble for the later stabilization. 
The scenario may also pose some complications for the 
understanding of  reheating
since the radion is long lived, and its mass could be very small (about $1/R$).

Another possible way out was proposed in Ref.~\cite{tye},
where it was suggested that the brane could be out of its stable point
at early times, and inflation is induced on the  brane by its movement
through the extra space. 
The common point of last  two scenarios is that they assume 
an unstable extra dimension along inflation process. 
This may also be troublesome as we shall discuss in next section

%%%%%%%%%%%%%%%%%%%%
\subsection{Cosmological Radion problem}

It is conceptually hard to accept the fact that the extra dimensions 
were at all times as large as suggested by ADD model. As it is natural to
think that all our observable Universe started as a small patch of size
$\sim M_\ast^{-1}$, it seems so for  the extra dimensions.
From its definition, an initial value $b_{in}=M_\ast^{-1}$ means 
that the radion started with a large negative value,
 \be 
   \label{init}
  \sigma_{in}~=~-M_{p}\sqrt{\frac{2(\delta+2)}{\delta}}
   \ln\left(\frac{M_{p}}{M_{\ast}} \right)~.
\ee
Since the stable point has been defined such that $\sigma|_{b=R}=0$, 
one has to conclude that the radion should roll down its potential from negative
values as the extra dimensions expand. 
If the radion potential were flat enough
in these range of values, it would be natural 
to think  that the radion could play
the role of the inflaton. The idea is enforced by the large
absolute initial value of $\sigma$ which may already satisfy chaotic initial
conditions for driving inflation. For realizing the picture, one has at some
point to be able to  calculate the radion potential from some fundamental
physics and proof that  it is indeed flat for negative $\sigma$ values, whereas
it  grows fast enough for positive $\sigma$'s to avoid dynamically 
driving the volume much beyond the expected stable value $b=R$. 
Without the actual potential is hard to make any serious calculation for
density perturbations or even reheating temperature. Certainly a simple mass
term like potential $U(\sigma)\sim m^2\sigma^2$ would not do it. Reason is two
fold. First, we are far away from the stable point where the radion mass is
defined,  so the potential hardly would be well described by 
just the second term of its Taylor expansion. Second, and more important, the
simple chaotic potential predicts insufficient density fluctuations for small
$M_\ast$. Indeed, the calculation gives~\cite{inf5d}
$\delta\rho/\rho\simeq m/M_P \ll M_\ast/M_P$.

Also interesting it is the possibility of producing inflation with 
the help of a bulk inflaton in presence of the radion field. 
This analysis can help us to understand whether our previous discussion 
makes at all sense. We now turn our attention
towards the action in Eq.~(\ref{sphi}). Notice that the inflaton to radion
coupling is given only through the inflaton potential. The coupling induces an
effective mass term for the radion which is proportional to Hubble scale, 
$m_{eff}^2\sim \alpha^2 V_{eff}(\phi)/M_P =  \alpha^2 H^2$. 
As a consequence, the radion gets a very steep 
effective potential term, which easily drives the radion towards 
the minimum within a Hubble time~\cite{mazumdar1}. 
Indeed, once Hubble induced mass term is switched on the radion
will follow the evolution
\be
\sigma(t)\sim \sigma_{in}~e^{-(m^2_{eff}/3H)t}\,,
\ee
where $\sigma_{in}$ is the initial amplitude.
Naively, one could think that
this should solve
the problem of stabilization, provided inflation last a bit longer
than this dynamical stabilization process.
Nevertheless, it has been realized that although this stabilization does take
place, it happens that the effective minimum of the potential does not coincide
with $\sigma=0$~\cite{kolb,mazumdar2}. Actually,  it generally happens that 
the  global minimum for $\sigma$ is  displaced 
during inflation when the radion potential 
is quite flat~\cite{kolb}. To see this, let us notice that the 
total potential has the form
 \be 
 U_{total}(\sigma,\phi)  = U(\sigma) +  e^{-\alpha \sigma/M_P}~ V_{eff}(\phi)~,
 \ee
where by definition, $U(\sigma)$ has a minimum at $\sigma=0$. 
During inflation $V_{eff}$ defines the Hubble scale, 
as already mentioned, which
is  taken as a constant. The global minimum for $\sigma$ of this potential 
is solution to the equation
 \be 
 U'(\sigma) - \frac{\alpha}{M_P}~e^{-\alpha \sigma/M_P} V_{eff} =0~.
 \ee
Clearly, $\sigma=0$ is not a global minimum. 
In fact, in order to match both terms of
the equation, 
the minimum should lay in the positive range
of $\sigma$ values. This implies that during inflation the extra space grows
beyond its stable size ($b=R$), and gradually comes back to the final stable
volume $V_\delta$ as the inflaton energy gets diminished~\cite{kolb,mazumdar2}.
How far we are from the expected value $V_\delta$
 depends on the actual profile of the radion potential. 
For steeper potentials it is easy to see  
the displacement could be negligible for practical proposes, 
however that is  not so for flatter
radion potentials~\cite{mazumdar2}, 
since a larger value on $\sigma$ would be  needed in the exponential of the 
last equation to match a small value of $U'(\sigma)$.
One interesting conclusion arises: inflation could in principle be 
consistently analyzed in the set up of an stable bulk, 
as we did in previous section, nevertheless
post-inflationary  effects of the radion dynamics have yet to be studied with
some care. 

%%%%%%%%%%%%%%%%%%%%%%%%%%%%%%%%%%%%%%%%%%%%%%%%%
\section{Non Factorizable Geometries: Randall-Sundrum Models.}
%%%%%%%%%%%
\subsection{Warped Extra Dimensions}

So far we have been working in the simplest picture where the energy density on
the brane does not affect the space time curvature, but rather it has
been taken as a perturbation on the flat extra space. 
However, for large brane densities this
may not be the case. 
The first approximation to the problem can be done by 
considering a five dimensional model where branes
are located at the two ends of a closed fifth dimension. 
Clearly, with a single extra dimension, the gravity flux produced by a single
brane at $y=0$ can not softly close into itself 
at the other end the space, making the model unstable, just as a charged
particle living in a closed one-dimensional world does not define a stable
configuration. Stability can only be
insured by the introduction of a second charge (brane).
Furthermore, to balance branes energy and still get 
flat (stable) brane metrics, one has to compensate the effect on the space by
the introduction of a
negative cosmological constant on the bulk. Hence, the fifth dimension would be 
a slice of an Anti de-Siter space with flat branes at its edges. 
Thus, one can keep the branes flat paying the
price of curving the extra dimension. Such curved extra dimensions are usually
referred as warped extra dimensions. 
Historically, the possibility was first 
mentioned by Rubakov and Shaposhnikov in Ref.~\cite{rubakov}, 
who suggested that the cosmological constant problem could be understood under
this light: the matter fields vacuum energy on the brane could be canceled
by the bulk vacuum, leaving a zero (or almost zero) cosmological constant for
the brane observer. No specific model was given there, though. 
It was actually  Gogberashvili~\cite{merab} who
provided the first exact solution for a warped metric, nevertheless, 
this models are best known after Randall and Sundrum (RS) 
who linked the solution to the 
the hierarchy problem~\cite{rs1}. Later developments suggested that the warped
metrics could even provide an alternative to compactification for the extra
dimensions~\cite{rs2,nimars}. In what follows we shall discuss a concrete
example as presented by Randall and Sundrum.

%%%%%%%%%%%%%%
\subsection{Randall-Sundrum background}

Lets consider the following setup. A five dimensional space with an
orbifolded fifth dimension of radius $r$ and coordinate $y$ which takes values
in the interval $[0,\pi r]$.
Consider two branes at the fixed (end) points
$y=0,\pi r$; with tensions $\tau$ and $-\tau$ respectively. 
For reasons that should become clear later on,  the brane at $y=0$
($y=\pi r$) is usually called the hidden (visible) or Planck (SM) brane. 
We will also assign to  the bulk  a negative cosmological constant $-\Lambda$. 
Contrary to our previous philosophy in ADD model, here we shall 
assume that all parameters are of the
order of the Planck scale. 
Next, we ask for the solution that gives a 
flat induced metric on the branes such that 4D Lorentz invariance is respected.
To get a consistent answer, one has to require that 
at every point along the fifth dimension the induced metric should be the
ordinary flat 4D Minkowski metric. Therefore, the components of the 5D metric
only depend on the fifth coordinate. Hence, one gets the ansatz
 \be 
  ds^2 = g_{AB} dx^A dx^B = 
  \omega^2(y)\eta_{\mu\nu}dx^\mu dx^\nu - dy^2~,
 \label{wmetric}
 \ee
where we parameterize $\omega(y) = e^{-\beta(y)}$.
The metric, of course, can always be written in different coordinate systems.
Particularly, notice that one can easily go to the conformally 
flat metric, where
there is an overall factor in front of all coordinates,  
$ds^2 =  \omega^2(z)[\eta_{\mu\nu}dx^\mu dx^\nu - dz^2]$, where the new
coordinate $z$ is a function of the old coordinate $y$ only.

Classical action  contains $S= S_{grav} + S_{h} + S_{v}$; where
 \be
 S_{grav} = \int \! d^4x\, dy \sqrt{g_{(5)}} \left( \frac{1}{2k_\ast^2} R_{5}
 +\Lambda\right) ~,
 \ee
gives the bulk contribution, whereas the visible and hidden brane 
parts are given by
 \be 
 S_{v,h}= \pm~\tau~\int\! d^4x \sqrt{-g_{v,h}}~,
 \ee
where $g_{v,h}$ stands for the induced metric at the visible and hidden
branes, respectively.

Five dimensional  Einstein equations for the given action, 
 \bea
 G_{MN} &=& R_{MN} - \frac{1}{2} g_{MN} R_{(5)} 
   \nonumber \\
 &=& -k_\ast^2\Lambda\,g_{MN} + 
   k_\ast^2\tau\, \sqrt{\frac{-g_{h}}{g_{(5)}}} 
    \delta_M^\mu \delta_N^\nu g_{\mu\nu}\delta(y) 
      \nonumber \\[1ex]
& &-  k_\ast^2\tau\, \sqrt{\frac{-g_{v}}{g_{(5)}}} 
    \delta_M^\mu \delta_N^\nu g_{\mu\nu}\delta(y-\pi r) 
\label{eers}
 \eea
are easily reduced into two simple independent equations. First, we can expand
the $G_{MN}$ tensor components on  last equation, using the metric
ansatz (\ref{wmetric}),  to show 
 \bea
 G_{\mu\nu} = -3\, g_{\mu\nu}\left( -\beta'' + 2 (\beta')^2 \right)~; \\
 G_{\mu 5} =0~; \quad \mbox{ and } \quad
 G_{55} = -6\, g_{55} (\beta')^2  ~.
 \eea
Next, using the RHS of Eq.~(\ref{eers}), one gets
 \be
 6 (\beta')^2 =  k_\ast^2  \Lambda~; 
 \ee
and 
\be 
 3\beta'' = k_\ast^2  \tau\, \left[\delta(y) -\delta(y-\pi r)\right]. 
\ee
Last  equation, clearly,  defines the boundary conditions for the function
$\beta'(y)$ at the two branes (Israel conditions). 
Clearly, the solution is  $\beta(y) = \mu |y|$, where 
 \be 
  \mu^2 =\frac{k_\ast^2\Lambda}{6}= \frac{\Lambda}{6 M_\ast^3}~,
 \label{rsmu}
 \ee
with the subsidiary fine tuning condition 
 \be 
 \Lambda = \frac{\tau^2}{6 M_\ast^3}~,
 \label{rsfine}
 \ee
obtained from the boundary conditions,  
that is equivalent to the exact cancellation of the effective four
dimensional cosmological constant. The background metric is therefore
 \be 
  ds^2 = e^{-2\mu |y|}\eta_{\mu\nu}dx^\mu dx^\nu - dy^2~.
 \label{rsmetric}
 \ee
The effective Planck scale in the theory is then given by 
 \be
 M_{P}^2 = \frac{M_\ast^3}{\mu}\left(1- e^{-2\mu r\pi}\right). 
 \label{rsmp}
 \ee
Notice that for large 
$r$, the exponential piece becomes negligible,
and above expression has the familiar form given in Eq.~(\ref{mp}) for one
extra dimension of (effective) size $1/\mu$.

%%%%%%%%%%%
\subsection{Visible versus Hidden Scales Hierarchy} 

The RS metric has a peculiar feature. Consider a given distance, $ds_0^2$,
defined by fixed intervals $dx_\mu dx^\mu$ from brane coordinates.
If one maps the interval from hidden to visible
brane, it would appear here 
exponentially smaller than what is measured at the
hidden brane, i.e., $ds_0^2|_{v} = \omega^2(\pi r) ds_0^2|_{h}$. 
This scaling property 
would have interesting consequences when introducing fields to live on any
of the branes. Particularly, let us discuss what happens for a theory 
defined on the visible brane.

The effect of RS background on visible brane field parameters 
is non trivial. Consider for instance the scalar field  
action for the visible brane at the end of the space  given by
 \[
 S_H =
\int\! d^4x\, \omega^4(\pi r) \left [\omega^{-2}(\pi r)
     \partial^\mu H\partial_\mu H - \lambda\left(H^2 - 
     \hat v_0^2\right)^2\right].
 \]  
As a rule, we choose all dimensionful parameters 
on the theory to be naturally given in terms of  $M_\ast$, and this 
to be close to $M_P$. 
So we take  $\hat v_0\sim M_\ast$.
After introducing the normalization 
$H\rightarrow \omega^{-1}(\pi r) H = e^{\mu r\pi} H$ to
recover  the canonical kinetic term, the above action becomes
 \be
 S_H = \int\! d^4x   
 \left[ \partial^\mu H\partial_\mu H - \lambda\left(H^2 - 
     v^2\right)^2\right],
 \ee
where the actual vacuum $v= e^{-\mu r\pi}\hat v_0$. Therefore, by choosing 
$\mu r \sim 12$, the physical mass of the scalar field, and its vacuum,
would naturally appear at the TeV scale rather than at the Planck scale, 
without the need of any large hierarchy on the radius~\cite{rs1}. 
Notice that, on the contrary, any field
located on the other brane will get a mass of the order of $M_\ast$. 
Moreover,  it also implies that no  particles exist in the visible brane with
masses larger than TeV. This observation has been consider a nice 
possible way of solving the scales hierarchy problem. For this reason, the
original model proposed that our observable Universe resides on the brane
located at the end of the space, the visible brane. So the other brane 
really becomes hidden. This two brane model is sometimes called RSI model.

%%%%%%%%%%%%%%%%
\subsection{Kaluza Klein decomposition}

As a further note, notice that since there is everywhere 4D Poincar\'e
invariance, every bulk field on the RS background can be expanded into four
dimensional plane waves $\phi(x,y)\propto e^{ip_\mu x^\mu} \varphi(y)$. This
would be  the basis for the Kaluza Klein decomposition, that we shall now
discuss.
Note also that the
physical four momentum of the particle at any position of the brane goes as 
$p_{phys}^\mu (y) = \omega^{-1}(y) p^\mu$. Therefore, modes which are soft on
the hidden brane, become harder at any other point of the bulk. 

Lets consider again a bulk scalar field, now on the RS background metric. The
action is then
 \be 
 S[\phi] = \frac{1}{2}\int\!d^4x\, dy \sqrt{g_{(5)}}
 \left(g^{MN}\partial_M\phi \partial_N\phi - m^2\phi^2 \right)~.
 \ee
By introducing the factorization $\phi(x,y) = e^{ip_\mu x^\mu} \varphi(y)$ 
into the equation of motion, one gets that the KK modes satisfy
 \be 
 \left[ - \partial_y^2 + 4\mu~ {\rm sgn}(y) \partial_y + m^2 +
 \omega^{-2}(y)\, p^2 \right] \varphi(y) =0~,
 \label{kkrs}
 \ee
where $p^2= p^\mu p_\mu$ can also be interpreted as the effective four
dimensional invariant mass, $m_n^2$. It is possible, through a functional  
re-parameterization and a change of variable, 
to show that
the solution for $\varphi$ can be written in terms of Bessel functions
of index $\nu = \sqrt{4 + m^2/\mu^2}$~\cite{gw0,Rubakov2}, as
follows 
 \be 
 \varphi_n(y) = \frac{1}{ N_n \omega^2(y)}\, 
 \left[ J_\nu \left( \frac{m_n}{\mu\omega(y)}\right) 
 + b_{n\nu} Y_\nu \left( \frac{m_n}{\mu\omega(y)}\right) \right]\,,
 \label{kkchi}
 \ee 
where $N_n$ is a normalization factor, $n$ labels the KK index, and the 
constant coefficient $b_{n\nu}$ has to be fixed by 
the continuity conditions at one of the boundaries. The other boundary
condition would serve to quantize the spectrum. For more details the reader
can see
Ref.~\cite{gw0}. Here we will just make some few comments about. 
First, for $\omega(\pi r)\ll 1$, 
the discretization condition that one gets for 
$x_{n\mu} = m_n/\mu\omega(y)$  looks as
 \be 
 2 J_\nu(x_{n\mu}) + x_{n\nu} J'_\nu(x_{n\mu}) =0~.
 \label{kkmrs}
 \ee
Therefore, the lowest mode satisfies $x_{1\mu}\sim {\cal O}(1)$, which means
that  $m_1 \simeq \mu e^{-\mu r \pi}$. For the same range of parameters we
considered before to solve the hierarchy problem, one gets that lightest KK
mode would have a mass of order TeV or so.
Next, for the special case of a originally massless field ($m=0$),
one has $\nu=2$, and thus the first solution to Eq.~(\ref{kkmrs}) is just
$x_{12}=0$, which indicates the existence of a massless mode in the spectrum.  
The next zero of the equation would be of order one again, thus the KK tower
would start at $\mu e^{-\mu r \pi}$. The spacing among two consecutive KK
levels would also be of about  same order. There is no need to stress that
this would actually be the case of the graviton spectrum.
This makes the whole spectrum
completely distinct from the former ADD model. With such heavy graviton modes 
one would not expect to have visible deviations on the short distance gravity
experiments, nor constrains from BBN or star cooling.

\subsection{Radion Stabilization}

The way RSI model solves the hierarchy problem between $m_{EW}$ and $M_P$ 
depends on the interbrane spacing $\pi r$. Stabilizing the bulk
becomes in this case an important issue if one is willing to keep this
solution. The dynamics of the extra dimension would give rise to a 
running away radion field, similarly  as it does for the ADD case.
A simple exploration of the metric (\ref{rsmetric}), 
by naively setting a time dependent bulk radius $r(t)$, shows that 
 \be 
 ds^2 \rightarrow e^{-2\mu r(t) |\phi|} \eta_{\mu\nu} dx^\mu dx^\nu - 
 r^2(t) d\theta^2~;
 \ee
with $\theta$ the angular coordinate on the half circle $[0,\pi]$. This suggest  
that if the interbrane distance changes the visible brane 
expands (or contracts) exponentially. 
The radion field associated to the fluctuations of the radius, 
$b(t) = r(t) - r$, is again massless
and thus it violates the equivalence principle.  
Moreover, without a stabilization mechanism for the radius, our brane
could expand forever. Some early discussions on this and other 
issues can be found in Refs.~\cite{gw,rsradion,rrs1}. 

The simplest and most elegant solution for stabilization in RSI 
was proposed by 
Goldberger and Wise~\cite{gw}. The central  idea is really simple: 
if there is a vacuum
energy on the bulk, whose configuration breaks translational invariance along
fifth dimension, say  $\langle E \rangle (y)$, then, the effective four
dimensional theory would contain a radius dependent potential energy
 \[ V(r) = \int\! dy\, \omega^4(y)\, \langle E \rangle (y) ~.\]
Clearly, if such a potential has a non trivial minimum, stabilization would be
insured. The radion would feel a force that tends to keep it at the
minimum.
The  vacuum energy $\langle E \rangle (y)$ may come from many sources.
The simplest possibility one could think up on is a vacuum induced by a bulk
scalar field, with non trivial boundary conditions, 
 \be 
 \langle \phi \rangle (0) = v_h \qquad \mbox{ and } \qquad 
 \langle \phi \rangle (\pi r) = v_v~.
 \ee
The boundary conditions would amount for a non trivial profile of 
$\langle \phi \rangle (y)$ along the bulk. Such boundary conditions may arise,
for instance, if $\phi$ has localized interaction terms on the branes, as
$\lambda_{h,v} (\phi^2 - v_{h,v}^2)^2$, which by themselves develop non zero
vacuum expectation values for $\phi$ located on the branes. 
The vacuum is then the $x$ independent solution 
to the equation of motion (\ref{kkrs}), which can be written as
 \be 
  \langle \phi \rangle (y) = \omega^{-1}(y)
  \left[ A \omega^{-\nu}(y) + B \omega^{\nu}(y)\right]~;
  \ee
where $A$ and $B$ are constants to be fixed by the boundary conditions.
One then obtains the effective 4D vacuum energy 
 \bea 
 V_\phi(r) &=&  \mu (\nu+2) A^2 \left(\omega^{-2\nu}(\pi r) - 1\right) 
 \nonumber \\
             &&    + \mu (\nu-2) B^2 \left(1 - \omega^{2\nu}(\pi r) \right)
 \eea
After a lengthly calculation, and in the limit where $m\ll \mu$
one finds that above potential has a non trivial minimum for 
 \be 
 \mu r = \left(\frac{4}{\pi}\right) \frac{\mu^2}{m^2} \ln
 \left[\frac{v_h}{v_v}\right]~.
 \ee
Hence, for $\ln(v_h/v_v)$ of order one, the stable value for the radius 
goes proportional to the curvature parameter, $\mu$, and inversely to the squared
mass of the scalar field. Thus, one only needs that $m^2/\mu^2\sim 10$ to get 
$\mu r\sim 10$, as needed for the RSI model.

One can get a bit suspicious about whether the vacuum energy 
$\langle \phi \rangle (y)$ may disturb the background metric. 
It actually does, although the correction is negligible as the 
calculations  for the Einstein-scalar field coupled equations 
may show~\cite{gw,rrs1}. 

%%%%%%%%%%%%%%%%%%%%%%%%%%%%%%%%%%%%%%%
\section{Infinite Extra Dimensions}

The background metric solution (\ref{rsmetric}) does not actually need 
the presence of the negative tension brane to hold as an exact solution 
to Einstein equations. Indeed the warp factor $\omega(y) = e^{-\mu |y|}$ 
has been  determined only by the Israel conditions at the $y=0$ boundary. That
is, by using   $\omega'' = \mu^2 \omega - \mu \omega \delta(y)$ in Einstein
equations, 
which implies equations (\ref{rsmu}) and (\ref{rsfine}). 
It is then tempting to `move' the 
negative tension brane to infinity, which  renders a non compact  fifth
dimension. The picture becomes esthetically more appealing, 
it has no need for compactification. 
Nevertheless, one has to ask now the question of whether such
a possibility is at all consistent with observations.  
It is clear that the Newton's constant is now simply
 \be 
 G_N = \mu G_\ast~
 \label{rsgn}
 \ee
--just take the limit $r\rightarrow \infty$ in Eq.~(\ref{rsmp})--, 
this reflects
the fact that although the extra dimension is infinite, 
gravity  remains  four dimensional at large distances (for $\mu r\gg 1$). 
This is, 
in other words, only a consequence of the flatness of the brane. 
We shall expand our discussion on this point in following sections.
Obviously, with this setup, usually called the RSII model, we are giving up the
possibility of explaining  the hierarchy between Planck and electroweak scales.
The interest on this model remains, however, due to potentially 
interesting physics at  low energy, and also due to  its connection to the
AdS/CFT correspondence~\cite{ads/cft}.

Although the fifth dimension is infinite, the point $y=\infty$ is in fact a
particle horizon. Indeed, the first indication comes from the metric, since 
$\omega(y\rightarrow\infty) = 0$. The confirmation would come from considering 
a particle moving away from the brane on the geodesics
$y_g(t) = \frac{1}{2\mu} \ln(1+\mu^2 t^2)$~\cite{ruba}. 
The particle accelerates towards
infinity, and its velocity tends to speed of light. The proper time interval is
then 
 \be 
 d\tau^2 = \omega^2(y_g(t)) dt^2 - \left(\frac{dy_g}{dt}\right)^2 dt^2~.
 \ee
Thus, the particle reaches infinity at infinite time $t$, but 
in a finite proper time $\tau = \pi/2\mu$.

\subsection{Graviton Localization}

In order to understand why gravity on the brane remains four dimensional at
large distances, even though the fifth dimension is non compact, one has to
consider again the KK decomposition for the graviton modes, with particular
interest on the shape for  the zero mode wave function. Consider first the
generic form of the perturbed background metric
 \[ ds^2 = \omega^2(y)g_{\mu\nu}dx^\mu dx^\nu + A_\mu dx^\mu dy - b^2 dy^2~.\]
Due to the orbifold projection $y\rightarrow -y$, the vector component $A_\mu$
has to be odd, and thus it does not contain a zero mode. Therefore at the zero
mode level only the true four dimensional graviton  and the scalar (radion)
should survive. Let us concentrate on the 4D graviton perturbations only.
Introducing the small field expansion as 
$g_{\mu\nu} = \eta_{\mu\nu} + \omega^{-2}h_{\mu\nu}$, and using the 
gauge fixing conditions $\partial_\mu h^\mu_\nu =0 = h^\mu_\mu$, 
one obtains  the wave equation 
 \be 
 \left[\partial_y^2  - 4\mu^2  - 
 \frac{m^2}{\omega^2(y)}  - 4\mu\delta(y)\right]h=0~;
 \ee
where the Lorentz indices should be understood. In the above equation the mass
$m^2$ stands for the effective four dimensional mass $p^\mu p_\mu = m^2$.
It should be noticed that the mass spectrum would now be continuous and 
starts at $m=0$. In this situation the KK are normalized to a delta function,
$\int\! dy~ \omega^{-2}(y)\, h_m(y)\, h_{m'} = \delta(m-m')$.
 
Introducing the functional re-parameterization
 \[ z = \frac{1}{\mu}sgn(y)\, \left(\omega^{-1}(y) -1\right)\]
and 
 \[ \Psi(z) = \omega^{-1/2}(y)~ h(y)~;\]
one can writes the equation of motion for the KK modes as the Scrh\"odinger
equation~\cite{rs2}
\be 
\left[ -\frac{1}{2} \partial_z^2 + V(z)\right]\Psi(z) = m^2 \Psi(z)
\ee
with a `volcano  potential' 
\be
V(z) = \frac{15\mu^2}{8(\mu |z| +1)^2} - \frac{3\mu}{2} \delta(z) ~,
\ee
which peaks as $|z|\rightarrow 0$ but has a negative 
singularity right at the origin.
It is well known from the quantum mechanics analog that such  
delta potential has a bound state, whose wave function is peaked at $z=0$, which
also means at $y=0$. In other words, there is a mode that 
appears as to be localized at the brane. 
Such a state is identified as our four dimensional graviton. Its localization is
the physical reason why gravity still behaves as four dimensional at the
brane.

Indeed, the wave function for the localized state goes as
 \be 
  \Psi_o(z) =\frac{1}{\mu (|z| +1/\mu)^{3/2}}
  \ee
whereas  KK mode wave functions in the continuum 
are written in terms of Bessel functions, in
close analogy to Eq.~(\ref{kkchi}), as
 \[ 
\Psi_m\sim s(z)
 \left[ Y_2\left( m|z| + \frac{1}{\mu}\right)
 + \frac{4\mu^2}{\pi m^2} J_2\left( m|z| + \frac{1}{\mu}\right) \right]
 \]
where $s(z) = (|z| +1/\mu)^{1/2}$. By properly normalizing these wave functions
using the asymptotics of the Bessel functions, 
it is possible to show that 
for $m<\mu$ the wave function at brane has the value
 \be 
 h_m(0) \approx \sqrt{\frac{m}{\mu}}~.
 \ee
The coupling of gravitons to the brane is therefore weak for the lightest
KK graviton states. The volcano potential acts as a barrier for those modes. 
The production of gravitons at low energies would then be negligible.

\subsection{Gravity  on the RSII brane}

The immediate application of our last calculations 
is on the estimation of the effective gravitational 
interaction law at the brane. The reader should remember that the 
effective interaction of brane matter to gravitons goes as
$h_{\mu\nu}(0)T^{\mu\nu}$. So, it involves the evaluation of the graviton wave
function at the brane position, as expected. Therefore the
graviton exchange  between  
two test particles on the brane separated by a distance $r$ 
gives the effective potential 
 \bea 
 U_{RSII}(r)&\approx& U_N(r)\left[1 + 
 \int_0^\infty\!\frac{dm}{\mu}\, \frac{m}{\mu} e^{-mr}\right]\\
 &=& U_N(r)\left[1 +\frac{1}{\mu^2r^2}\right]~.
 \eea
Notice that the correction looks exactly as in the two extra dimensional ADD
case, with  $1/\mu$ as the effective size of the extra
dimensions. Thus, to the brane  the bulk should appear as compact,
at least from the gravitational point of view. The conclusion is striking.
There could be non compact extra dimensions and yet scape to our observations!.

%%%%%%%%%%
\subsection{Higher dimensional generalization}

The RSII model, which  provides a serious alternative to compactification, 
can immediately be extended to have a larger number of dimensions. 
First, notice that the metric (\ref{rsmetric}) has came from the peculiar
properties of co-dimension one objects in gravity. Thus, it is obvious that the
straightforward generalization should also contain some co-dimension one branes
in the configuration. Our brane, however should have a larger co-dimension.
Lets consider a system of $\delta$ mutually intersecting 
$(2+\delta)$ branes in a $(4+\delta)$
dimensional AdS space, of cosmological constant $-\Lambda$. 
All branes should have a positive tension $\tau$.
Clearly, the branes intersection is a 4 dimensional 
brane, where we assume our Universe lives. 
Intuitively, each of the $(2+\delta)$
branes would try to localize the graviton to itself, just as the RSII brane
does. As a consequence, the zero mode graviton would be localized at the
intersection of all branes. This naive observation can indeed be confirmed by
solving the Einstein equations for the action~\cite{nimars}
 \bea 
 S &=& \int\!d^4x\, d^\delta y\, \sqrt{|g_{(4+\delta)}|} 
 \left( \frac{1}{2k_\ast} R_{(4+\delta)}
 +\Lambda\right) 
 \nonumber \\
 &&-\sum_{\mbox{all branes}} \tau\,\int\!d^4x \, d^{\delta-1}y\,
 \sqrt{|g_{(3+\delta)}|} 
 \eea
If the branes are all orthogonal to each other, 
it is straightforward to see that  the space consist of $2^\delta$ equivalent 
slices of AdS space, glued together along the flat branes. 
The metric, therefore,  would be conformally flat. Thus, one can write it down
using appropriate bulk coordinates as
 \be   
 ds_{(4+\delta)}^2 = \Omega(z) \left(\eta_{\mu\nu} dx^\mu dx^\nu - 
 \delta_{kl}\, dz^k dz^l\right)
 \label{nrsmetric}
 \ee
with the warp factor 
 \be 
 \Omega(z) = \frac{1}{\mu \sum_j |z^i| +1}
 \ee
where the $\mu$ curvature parameter is now
 \be 
 \mu^2 = \frac{2 k_\ast^2 \Lambda}{\delta(\delta+2)(\delta+3)};
 \label{nrsmu}
 \ee
which is a generalization of the relation given in Eq.~(\ref{rsmu}).
Similarly, the fine tuning condition (\ref{rsfine}), now  looks as
 \be 
 \Lambda = \frac{\delta(\delta+3)}{8(\delta+2)}\tau^2k_\ast^2~.
 \label{nrsfine}
 \ee
Effective Planck scale is now calculated to be 
 \be 
 M_P^2 = M_\ast^{(\delta+2)}\int\!d^\delta z\Omega^{(2+\delta)}= 
 \frac{2^\delta\delta^{\delta/2}}{(\delta+1)!} M_\ast^{(\delta+2)} L^\delta~,
 \ee
for $L= 1/\sqrt{\delta}\mu$. Notice this expression resembles the ADD relationship
given in Eq.~(\ref{mp}), with $L$ as the effective size of the extra dimensions.

Graviton localization can be now seen by perturbing the metric with 
$\eta_{\mu\nu}\rightarrow \eta_{\mu\nu} +h_{\mu\nu}$ in Eq.~(\ref{nrsmetric}),
and writing down the equation of motion for $h_{\mu\nu}$ in the gauge
${h^\mu}_\mu=0=\partial_\mu h^{\mu\nu}$, and in conformal coordinates, to get
for $\Psi= \Omega^{(\delta+2)/2} h$ the linearized equation
 \be 
 \left[ -\frac{1}{2} m^2 +\left( -\frac{1}{2} \nabla_{z}^2 +
 V(z)\right)\right] \hat \Psi =0~,
 \ee
which is again nothing but a Schr\"odinger equation 
with  the effective potential 
 \be 
  V(z) = \frac{\delta(\delta+2)(\delta+4) \mu^2}{8}\Omega -
  \frac{(\delta+2)\mu}{2}\Omega\sum_j \delta(z^j)
  \ee
Indeed, the spectrum has a massless  bound state localized around 
the intersection of all delta function potentials ($z=0$), which goes as
$\Psi_{bound}\sim \Omega^{(\delta+2)/2}(z)$. Since the potential falls off to zero at
large $z$, there would also be a continuum of modes. Since the height of the
potential near the origin goes as $\mu^2$, all modes with small  masses, $m<\mu$
will have suppressed wave functions, while those with large masses will be
unsuppressed at the origin. Therefore, the contribution of the lightest modes to
gravitational potential for two test particles at the brane 
would again come suppressed as in the RSII case. The correction to Newton's law 
goes as~\cite{nimars}
 \be
 \Delta U(r)\sim U_N(r)\left( \frac{L}{r}\right)^\delta ~,
 \ee
which again  behaves as in the ADD case, mimicking the case of compact
dimensions, thought there are not so.

%%%%%%%%%%%%%%%%%%%%%%%%%%%%%%%%%%%%%%%%%%%%%%%%%%%%%%%%
\section{Brane Cosmology}

Let us now discuss what modifications are introduced to cosmology if one
considers the RSII setup. One of the first things that one
needs to know is the time
dependence of the metric. 
As the bulk curvature arose to compensate the brane 
tension in order to keep the brane flat, once a time dependent 
energy density $T_{\mu\nu}$ 
is introduced on the brane, as needed for our Universe, 
also the warping will become time dependent, and so, one has to reconsider the
RS solution to five dimensional Einstein equations. 
The problem was first
addressed by Binetruy et al. in Refs.~\cite{binetruy,binetruy2}. 
It has also been noted that Friedmann
equation could be recovered on the brane in the low energy limit
($\rho\ll\tau$),
on the basis of the  fine tuning  (\ref{rsfine}), 
even thought the metric is  not
static~\cite{waldram,cline,cosmo5d}. 
Here we will discuss some of these results.

We  start by considering what the metric ansatz should be for brane cosmology,
assuming that the only energy sources are the brane energy momentum tensor,
$T_{\mu\nu}$, and the negative cosmological constant of the brane. 
We adopt the cosmological principle of isotropy and homogeneity 
in the three space
dimensions of the brane, thus,  the most general 
$T_{\mu\nu}$ has a diagonal form, parameterized by energy density, $\rho$, 
and pressure, $P$, 
\[ {T^\mu}_\nu = diag(\rho, -P, -P,-P)~.\]
Also, this implies that  $g_{MN} = g_{MN}(y,t)$, only, where
the $y$ dependence reflects the breaking of translational 
invariance along fifth dimension due to the presence of the brane.
Finally, and for simplicity, we will consider that distance intervals along 
fifth dimensions are fixed along time.
We then chose the following ansatz for the metric
 \be
  ds^2=w^2(y,t)\, dt^2 - a^2(y,t)\,\gamma_{ij}\,dx^i dx^j -  dy^2~,
  \label{bcmetric}
 \ee
were $\gamma_{ij} = f(r) \delta_{ij}$, with $f^{-1}(r)=1-kr^2$ being the
usual Robertson-Walker curvature term, where $k=-1,0,1$. For simplicity, we
shall restrict our discussion to the flat Universe case ($k=0$). 
It is worth noticing  that  such a 
metric does reduce to the standard Friedmann-Robertson-Walker metric 
(\ref{frw}) when evaluating
distance intervals at the brane, provided $w(0,t) = 1$. 
As in RS models we will here assume that the 
orbifold symmetry  $P: y\rightarrow -y$ is present, thus, $a$ and $w$ 
depend only on $|y|$.

The five dimensional Einstein equations take the form
 \be
 G_{AB} = R_{AB}-\frac{1}{2} g_{AB} R = 
 k_\ast^2 \left[-\Lambda g_{AB} + 
 T_{\mu\nu}\ \delta^\mu_A\ \delta^\nu_B\ \delta(y)\ \right],
 \label{ecosmo}
 \ee
The delta function on the RHS of Eq.~(\ref{ecosmo}) can be understood as a
boundary condition. We then proceed by solving first the equations away from the
brane. The global metric solution clearly has to be continuous everywhere. That
is naturally solved by the orbifold condition.  Metric derivatives on $y$,
however, are not continuous, they should have a gap at $y=0$, which should match
the brane energy momentum tensor,
 \be 
  \int^{0+}_{0^-} \! dy~ G_{\mu\nu} = k_\ast^2 T_{\mu\nu}.
  \label{int}
 \ee     
Let us now proceed to the details.  We use the metric ansatz 
(\ref{bcmetric}) to explicitely expand RHS of Eq.~(\ref{ecosmo}) to
get~\cite{binetruy}
 \bea
  G_{00}&=& 
   3\left\{ \left( \frac{\dot a}{a}\right)^2 -
   w^2 \left[ \left(\frac{a'}{a} \right)^2  + \frac{a''}{a}\right]  \right\}
   \nonumber\\
  G_{ij}&=& \left[
   a^2 \left\{ \frac{a'}{a} \left( 2\frac{w'}{w} + 
   \frac{a'}{a} \right) + 2\frac{a''}{a} + 
   \frac{w''}{w}\right\} \right.\nonumber \\
   && \left. + \frac{a^2}{w^2} \left\{\frac{\dot a}{a} 
   \left(2\frac{\dot w}{w}-\frac{\dot a}{a}\right)  
   - 2\frac{\ddot a}{a} \right\}\right]\delta_{ij} ~,
    \label{Gij}\\
  G_{04}&=&
   3\left(\frac{w'}{w}\frac{\dot a}{a}
   -\frac{\dot a'}{a}\right), 
   \nonumber\\
  G_{44}&=&
   3\left\{\frac{a' }{a}\left(\frac{a' }{a}+\frac{w' }{w}\right)-
   \frac{1}{w^2}\left[\frac{\dot a}{a} \left(\frac{\dot a}{a}-
   \frac{\dot w}{w}\right) + \frac{\ddot a}{a}\right] \right\}
   \label{G44} 
 \nonumber
 \eea 
where primes  denote derivatives with respect to $y$ and dots derivatives with
respect to $t$. Our boundary (Israel) conditions are then
\bea
  \frac{\Delta a'}{a }\bigg|_{y=0} &=& -\frac{k_\ast^2}{3}\rho, 
  \label{gap1}\\
  \frac{\Delta w'}{w }\bigg|_{y=0} &=& \frac{k_\ast^2}{3}(3P + 2\rho);
   \label{gap2}
 \eea 
here  the gap function  $\Delta a'(0)= a'(0^+) - a'(0^-) = 2 a'(0^+)$ 
gives  the size of the jump for the $y$ derivative of $a$ at zero. 
The same applies for $\Delta w'$. We can straightforwardly use these 
boundary conditions to evaluate the Einstein tensor components at the brane.
Particularly, equation for $G_{04}$ gives the energy conservation formula
 \[ \dot \rho + 3H_0\left( \rho +P\right) =0 \]
where we have introduced the brane Hubble function, $H_0(t)$,  as 
the $y=0$ value of the more general bulk Hubble function~\cite{cosmo5d} 
 \be 
 H(y,t)\equiv \left(\frac{\dot a}{a}\right)~.
 \ee  
It is also not difficult to show that the equation $G_{04}=0$
implies to more  general condition 
 \be 
 w(y,t) = \lambda(t)\dot a(y,t)~.
 \label{n}
 \ee

It is very illustrative to rewrite the equation for $G_{00}$  as 
bulk Friedmann equation~\cite{cosmo5d}
  \be 
  H^2(y,t) = w^2 \left[ - \frac{k_\ast^2}{3} \Lambda  + 
  \left(\frac{a'}{a}\right)^2 + \frac{a''_R}{a}\right] ~;
  \label{rsH}
 \ee  
where $a''_R$ stands for the regular part of the function.
It is then clear that upon evaluation at the brane, the Friedmann equation
presents the `wrong' dependence  $H_0^2\propto \rho^2$~\cite{binetruy} coming
from the second term on the RHS of Eq.~(\ref{rsH}). 
Also, we may identify the last term of same equation as the
contribution of the Weyl tensor of the bulk~\cite{shiru}. 
We will come back to this point a bit later. Now we turn into the other 
equations. First, $G_{ij}$ gives a non independent equation. Indeed, 
it can  be derived  from Eq.~(\ref{rsH}) by taking a time derivative, and 
combining the result  with $G_{04}$. This is same as in the usual 4D case where
$G_{ij}$ gives the acceleration equation. Finally, $G_{44}$ 
represents the only truly new equation in the system. It is also 
the window to solve the $y$ dependence of the metric, since combined with the
bulk Friedmann equation (\ref{rsH}) and the acceleration equation, 
it simplifies to
 \be
 3\frac{a''}{a}+\frac{w''}{w}= \frac{2}{3}k_\ast^2\Lambda~.
 \label{a''}
 \ee
Solving this equation, together with (\ref{n}),  we get  the  result
 \bea
 a(y,t) = a_0(t)\, \left( \cosh(\mu|y|) - 
 \frac{k_\ast^2}{6\mu}\, \rho\, \sinh(\mu|y|)\right), \nonumber\\
 w(y,t) =  \cosh(\mu|y|) +
  \frac{k_\ast^2}{6\mu}\,(3P + 2\rho)\,\sinh(\mu|y|); 
  \label{nasol}
 \eea
where  the bulk curvature parameter
$\mu^2 = k_\ast^2\Lambda/6$. Notice that in the static limit $(a_0 =1)$
where $\rho=-P=\tau$, with the brane tension obeying the fine tuning condition
(\ref{rsfine}), we recover the RS metric solution $a(y)=w(y) = e^{-\mu|y|}$.
Notice also that in the time dependent case 
we do get the FRW metric on the brane.

We can now completely evaluate the Friedmann equation (\ref{rsH}) 
on the brane by  using that the total energy density 
$\rho = \rho_m + \tau$, 
with $\rho_m$ the actual brane matter density, to get~\cite{cline}
 \be 
 H^2_0(t) = \frac{\rho_m}{3M_P} \,\left( 1+ \frac{\rho_m}{2\tau}\right),
 \label{Hbrane}
 \ee
which have a quadratic term on the matter density.
A Universe described by such a modified Friedmann equation 
evolves faster than the standard one. This may not be a problem during
the very early stages of the Universe, 
whereas just before Nucleosynthesis  the standard cosmological
evolution $H_0^2\sim \rho_m$ has to had be restored to not disturb the success 
of the theory.  
Clearly, for small matter densities $\rho_m\ll \tau$
one recovers the standard Hubble expansion. 

It is interesting to note, on the other hand, that inflation  
when drove by a scalar field  whose energy density 
over passes the brane tension is more efficient in the brane world. 
This can be seen from the equation of motion 
$\ddot \varphi + 3H_0\varphi + V'(\varphi) =0$ where the 
friction term becomes larger for  larger  energy densities.
Thus slow roll gets enhanced by the modification to the Friedmann equation
(\ref{Hbrane}).
Inflation would then last longer than in the standard 4D models, 
and even some steep potentials that were unable to drive inflation in the 4D
case, could now be successful~\cite{maartens}. Expansion at high energies 
drives the tilt of the spectrum of adiabatic 
density perturbations to zero and it seems not to 
alter their expected amplitude~\cite{maartens}. 
Physics of reheating~\cite{copeland,amp}, preheating, 
and other pre-BBN phenomena may also be affected, 
depending on the energy scale at which they take place.

\subsection{Geometric approach}

A more formal and general 
treatment of the brane model for the derivation of the effective
Einstein equations on the brane was presented in Ref.~\cite{shiru}. 
It   uses a covariant geometric approach that does not  rely 
on the metric ansatz, and I believe,  it is worth  to be underlined in here.
Let us denote the unit vector normal  to the brane  by $n^A$, 
and the induced brane metric as $\gamma_{AB} =  g_{AB} - n_A n_B$.
Next we consider the extrinsic curvature of the brane 
 $K_{AB} = \gamma_A^C\,\gamma_B^D \nabla_C n_D $
to write down the  Gauss--Codacci equations
 \be
 {R_{(4)}}^A_{BCD} = {R_{(5)}}^M_{NPQ}\,
 \gamma^A_M\,\gamma^N_B\,\gamma^P_C\,\gamma^Q_D +  
 2K^A_{[C} K_{B]D}~,
 \ee
 \be
 D_N K^N_M  - D_M K = {R_{(5)}}_{PQ} n^Q \gamma^P_M~,
 \ee
where $D_M$ is the covariant differentiation with respect to $\gamma_{MN}$. 
The brane Ricci tensor is  then obtained by contracting the first
of above equations on $A$ and $C$, and one gets 
${R_{(4)}}_{MN}= {R_{(4)}}_{CD}\,\gamma^C_M\,\gamma^D_N - 
{R_{(4)}}^A_{BCD}n_A\,\gamma^B_M\, n^C\gamma^D_N + K K_{MN} -  K^A_M K_{NA}$. 
A further contraction on $M$ and $N$ will provide us the scalar curvature
$R_{(4)}$. All together shall define  
the five dimensional Einstein equation with a source  given by 
 \be 
T_{MN} = \Lambda g_{MN} + S_{MN}\delta(\chi)~,
\ee
where we have explicitely chosen $\chi$ as 
the locally orthogonal coordinate to the brane, without lost of generality.
Here $S_{MN}$ represents the brane energy density, which goes as
 \be 
 S_{MN}=  -\tau\, \gamma_{MN} + t_{MN}
 \label{s}
 \ee
with $t_{MN}$ the brane energy momentum tensor, which clearly satisfies
$t_{MN}n^M=0$. It should be noted that 
properly speaking $S_{MN}$ should be evaluated by the variational
principle of the 4D Lagrangian for matter fields. The decomposition (\ref{s})
can be ambiguous. 
Again, the delta function would lead us to the Israel junction
conditions
 \be 
 [\gamma_{MN}] = 0\quad \mbox{ and } 
 \quad [K_{MN}]= -k_\ast^2\left( S_{MN} - \frac{1}{3} \gamma_{MN} S\right)~, 
 \ee
where $[X]\equiv \Delta X(0)= 2X(0^+)$, due to the $Z_2$ symmetry.
First of this expressions only states the continuity of the metric at the brane,
whereas the other allow us to completely 
determine the extrinsic curvature of the brane in terms of the energy momentum
tensor.
Putting all those equations together, 
one gets the effective brane Einstein equations 
 \be 
 {G_{(4)}}_{MN} = \Lambda_4 \gamma_{MN} + 8\pi G_N t_{MN} 
 + k_\ast^2 \pi_{MN} -  E_{MN}~,
 \ee
where the effective brane cosmological constant 
 \be 
 \Lambda_4 = \frac{1}{2} k_\ast^2
  \left( \frac{1}{6} k_\ast^2 \tau - \Lambda\right)
 \ee
is null only if the fine tuning condition (\ref{rsfine}) holds.
Newton constant is defined in terms of the brane tension  by
 \be
 8\pi G_N = \frac{k_\ast^2 \tau}{6}~,
 \ee
expression that is equivalent to Eq.~(\ref{rsgn}), with the proper
insertion of the parameter $\mu$ as defined in Eq.~(\ref{rsmu}).
$E_{MN} $ stands for the limit value at $\chi=0^+$ of the 
5D  Weyl tensor  ${C^{(5)}}_{MANB}n^An^B$, 
with ${C^{(5)}}$ the 5D Weyl curvature tensor. This gives the non
local effects from the free gravitational field in the bulk and it cancels when
the bulk is purely AdS. 
And last but not least, the tensor $\pi_{MN}$ gives a local  quadratic
contribution from  the  brane energy momentum tensor which arises
from the extrinsic curvature terms. It goes  as
 \be 
 \pi_{MN} = \frac{1}{12} t\, t_{MN} -\frac{1}{4} t_{MA}t^A_M 
 + \frac{1}{24} \gamma_{MN}\left[ 3t_{AB}t^{AB} - t^2\right].
 \ee

%%%%%%%%%%%%%%%%%%%%%%%%
\section{Concluding remarks}

Along the present notes I  have introduced the reader to some 
aspects of models with extra dimensions where our Universe is constrained 
to live on a four dimensional hypersurface. The study of brane world
has become a fruitful industry that has involved several areas of 
theoretical physics in matter of  few years. 
It is fair to say, however, that many of the current leading directions 
of research  obey more to speculative ideas that to well established facts. 
Nevertheless, as it happens with any other  physics speculation, 
the studies on the brane world are
guided by the principle of physical and mathematical 
consistency, and the further  possibility of 
connecting the models with  more fundamental theories, i.e. String
theory, from where the idea of extra dimensions and branes had been borrowed;
and so with the experiment in the near future.

It is hard to address the too many interesting  
topics of the area with the detail I intended in here, 
without facing  trouble with the space of this  short notes. 
In exchange, I have concentrated the discussion to the construction of  
the main  frameworks (ADD, and RS models), 
the calculation of the effective 
gravity interactions on the brane, and brane cosmology, mainly.
I hope these could serve the propose of introducing the reader 
to this area of research.
The list of the left behind is extensive, it includes the 
recent discussions on the cosmological constant problem~\cite{cc},
higher dimensional warped spaces~\cite{hdw} 
dark matter from KK modes~\cite{dmkk}, Black Holes in both ADD and RS
models~\cite{bh1,rsbh}, 
deconstruction of extra dimensions~\cite{deconstruction}, and the list go on.
I urge the further interested readers to go to the more extended
reviews~\cite{reviews} and  hunt for further references.

\end{document}